\documentstyle[epsf]{article}    
\def\title#1{\begin{centering}\Large\bf #1 \\[5mm]\end{centering}}
\def\author#1{\centerline{#1}}
\def\address#1{\begin{centering}\bf #1 \\[5mm]\end{centering}}

\newcommand{\be}{\begin{equation}}
\newcommand{\ee}{\end{equation}}
\newcommand{\vp}{\varphi}
\newcommand{\br}{{\bf r}}
\newcommand{\dlt}{\delta}
\newcommand{\om}{\omega}
\newcommand{\Dlt}{\Delta}
\newcommand{\prt}{\partial}
\newcommand{\al}{\alpha}
\newcommand{\bt}{\beta}
\newcommand{\ra}{\rightarrow}
\newcommand{\gm}{\gamma}
\newcommand{\Lbd}{\Lambda}

\begin{document}
\small
%
%
\title{Topological Coherent Modes in Trapped Bose Gas}
\author{{\large{\bf V.I. Yukalov$^{1,2}$, K.-P. Marzlin$^3$, 
E.P. Yukalova$^{1,4}$,}}}
\author{{\large{\bf and V.S. Bagnato$^5$}}}
\vskip 5mm

\address{$^1$Institut f\"ur Theoretische Physik, \\
Freie Universit\"at Berlin, Arnimallee 14, D-14195 Berlin, Germany}
\address{$^2$Bogolubov Laboratory of Theoretical Physics, \\
Joint Institute for Nuclear Research, Dubna 141980, Russia}
\address{$^3$Institute for Quantum Information Science, \\
2500 University Drive NW, Calgary, Alberta T2N 1N4, Canada}
\address{$^4$Department of Computational Physics, \\
Laboratory of Information Technologies,\\
Joint Institute for Nuclear Research, Dubna 141980, Russia}
\address{$^5$Research Center for Optics and Photonics, \\
Instituto de Fisica de S\~ao Carlos, \\ 
University de S\~ao Paulo, S\~ao Carlos, S\~ao Paulo 13560-970, Brazil}

\vskip10.pt

\begin{abstract}

The report reviews the problem of topological coherent modes, which are  
nonlinear collective states of Bose-condensed atoms. Such modes can be 
generated by means of alternating external fields, whose frequencies are 
in resonance with the transition frequencies between the related modes. 
The Bose gas with generated topological coherent modes is a collective 
nonlinear analog of a resonant atom. Such systems exhibit a variety of 
nontrivial effects, e.g.: interference fringes, interference current, 
mode locking, dynamic transitions, critical phenomena, chaotic motion,
harmonic generation, parametric conversion, atomic squeezing, and 
entanglement production.

\end{abstract}

\vskip10.pt

\section{Introduction}

Trapped Bose gas, being dilute and cooled down to temperatures below the
Bose-Einstein condensation temperature, forms a coherent system well described 
by the Gross-Pitaevskii equation [1--5]. Stationary solutions to the latter are 
called the topological coherent modes [6]. The ground state, corresponding to 
the lowest energy level of the Gross-Pitaevskii stationary equation, represents 
the standard Bose-Einstein condensate. The states, associated with the higher 
energy levels, describe nonground-state condensates. The higher topological 
coherent modes can be generated by means of alternating fields, with the 
frequencies that are in resonance with the chosen transition frequencies [6].

The name {\it topological coherent modes} comes from the following. Different
stationary solutions to the Gross-Pitaevskii equation, related to distinct
energy levels, possess essentially dissimilar spatial shapes, with different
number of zeros. The modes, displaying distinct spatial topology, can be named
{\it topological}. These modes can be characterized as {\it coherent} due to the
fact that the Gross-Pitaevskii equation can be interpreted as an exact equation
for coherent states [7].

The topological coherent modes, described by the nonlinear Gross-Pitaevskii
equation, should not be confused with elementary collective excitations,
defined by the linear Bogolubov - de Gennes equations. The elementary
excitations describe small oscillations around a given nonlinear topological
mode and do not change the topology of the latter [1].

The general notion of the nonlinear coherent modes and the possibility of their
generation by means of resonant fields was advanced in Ref. [6]. A particular
case of vortex creation was considered in Refs. [8,9]. Various properties of
topological coherent modes were also studied in theoretical papers [10--27]. A
dipole topological mode was generated in experiment [28]. The feasibility of
resonant formation of multimode condensates, consisting of several topological
coherent modes, was investigated [29,30].

Bose gas of trapped atoms with resonantly generated topological coherent modes
exhibits a variety of unusual features. The aim of this report is to present a
general picture of such a resonant system (Section 2) and to give a survey of
its most interesting properties (Section 3).

\section{Topological Coherent Modes}

The topological coherent modes are defined [6] as the stationary solutions to 
the eigenvalue problem
\be
\label{1}
\hat H[\vp_n]\; \vp_n(\br) = E_n\; \vp_n(\br) \; ,
\ee
which is represented by the stationary Gross-Pitaevskii equation with the 
nonlinear Hamiltonian
\be
\label{2}
\hat H[\vp] \equiv -\; \frac{\hbar^2}{2m}\; \nabla^2 +U(\br) +
N A_s|\vp|^2 \; ,
\ee
containing a trapping potential $U(\br)$ and the interaction intensity
$A_s\equiv 4\pi\hbar^2 a_s/m$, with $m$ being atomic mass; $a_s$ scattering 
length; and $N$ the total number of atoms.

In what follows, we shall use the notation for the scalar product
$(\vp_m,\vp_n)\equiv\int\vp_m^*(\br)\vp_n(\br)d\br$.
The eigenfunctions $\vp_n(\br)$, because of the nonlinearity of problem (1), 
are not necessarily orthogonal, so that $(\vp_m,\vp_n)$ is not compulsory the 
Kroneker delta $\dlt_{mn}$. But the functions $\vp_n(\br)$ can always be 
normalized, with $(\vp_n,\vp_n)\equiv||\vp_n||^2=1$.

If the trapped Bose gas is initially in one of the coherent modes $n$, then to
generate another mode requires applying an alternating field
$$
V(\br,t)= V_1(\br)\cos\om t + V_2(\br)\sin \om t \; ,
$$
with a frequency $\om$ being close to one of the transition frequencies
\be
\label{3}
\om_{mn} \equiv \frac{1}{\hbar}\; (E_m -E_n) \; .
\ee
Say the modes with the energies $E_1$ and $E_2$ are connected, such that 
$E_1<E_2$, with the transition frequency being $\om_{21}$. The related 
resonance condition reads as
\be
\label{4}
\left | \frac{\Dlt\om}{\om} \right | \ll 1 \; , \qquad
\Dlt\om\equiv \om -\om_{21} \; .
\ee

With an applied time-dependent field, we have the temporal Gross-Pitaevskii
equation
\be
\label{5}
i\hbar\; \frac{\prt}{\prt t} \; \vp(\br,t) = \left ( \hat H[\vp] +
\hat V\right ) \; \vp(\br,t) \; ,
\ee
in which $\hat V=\hat V(\br,t)$. The alternating field can be represented as
\be
\label{6}
V(\br,t) = \frac{1}{2}\; B(\br)\; e^{i\om t} + 
\frac{1}{2}\; B^*(\br)\; e^{-i\om t} \; ,
\ee
where $B(\br)\equiv V_1(\br)-iV_2(\br)$.
Experimentally, the alternating field can be realized as the modulation of the
trapping potential. Another way could be by alternating the scattering length
$a_s(t)$ by modulating an external magnetic field close to a Feshbach resonance
[31--33].

One possibility of studying the resonant generation of coherent modes would 
be by a direct numerical solution of the temporal equation (5), which we have 
done in our works [29,30]. However to get a deep physical insight into the 
problem, it is necessary to develop an analytical theory. Such a general 
theory, based on the averaging technique [34], was developed in Refs. 
[6,12,15,20--24,29,30]. In order to show that the basic equations of the 
analytical approach can be accurately derived and all consideration is well 
mathematically grounded, we describe, first of all, the main steps of this 
derivation.

We can look for the solution of Eq. (5) in the form of the mode expansion
\be
\label{7}
\vp(\br,t) = \sum_n c_n(t) \vp_n(\br) \exp\left ( -\;
\frac{i}{\hbar}\; E_n t\right ) \; ,
\ee
with the coefficients $c_n(t)$ being slow functions of time, such that
\be
\label{8}
\frac{\hbar}{E_n}\; \left | \frac{dc_n}{dt} \right | \ll 1 \; .
\ee
Then the functions $c_n(t)$ can be treated as temporal quasi-invariants with
respect to the fast exponentials $\exp(-iE_nt/\hbar)$. Substituting expansion
(7) into Eq. (5), we multiply the latter by the mentioned exponential and 
average over time according to the rule
$$
\lim_{T\ra\infty} \; \frac{1}{T}\; \int_0^T f(c_n,t)\; dt \; ,
$$
where the quasi-invariants $c_n$ are kept fixed. Averaging in this way the
equality $||\vp||^2=1$, we have the normalization condition
\be
\label{9}
\sum_n |c_n(t)|^2  = 1\; .
\ee

For what follows, we need the notation for the transition amplitudes: one of 
them being due to atomic interactions,
\be
\label{10}
\al_{mn} \equiv A_s\; \frac{N}{\hbar} \; \left ( |\vp_m|^2,\; 
2|\vp_n|^2-|\vp_m|^2 \right ) \; ,
\ee
and another related to the external field (6),
\be
\label{11}
\bt_{mn} \equiv \frac{1}{\hbar}\left (\vp_m,\hat B\vp_n\right ) \; ,
\ee
where $\hat B\equiv B(\br)$. Note that from normalization (9) one has
$|c_n|^2=1-\sum_{m(\neq n)}|c_m|^2$. Using this, the nonlinear part, 
resulting from Eq. (5), can be represented as
$$
A_s\; \frac{N}{\hbar}\; \sum_m (2-\dlt_{mn})(|\vp_n|^2,|\vp_m|^2) =
\sum_{m(\neq n)}\; \al_{nm}|c_m|^2 c_n \; + \al_{nn} c_n \; .
$$
Thus we come to the equation
\be
\label{12}
i\; \frac{dc_n}{dt} = \sum_{m(\neq n)} \al_{nm} |c_m|^2 c_n \; + \al_{nn} c_n +
\frac{1}{2}\; \dlt_{n1}\bt_{12} c_2 e^{i\Dlt\om \cdot t} +
\frac{1}{2}\; \dlt_{n2}\bt_{12}^* c_1 e^{-i\Dlt\om\cdot t} \; .
\ee

What is of physical interest is the behaviour of the fractional mode populations
\be
\label{13}
w_n(t) \equiv |c_n(t)|^2 \; .
\ee
These do not depend on the phase of $c_n$. Therefore, we may employ the gauge
transformation
\be
\label{14}
c_n \longrightarrow c_n\exp(-i\al_{nn} t) \; .
\ee
Then Eq. (12) reduces to
\be
\label{15}
i\; \frac{dc_n}{dt} = \sum_{m(\neq n)} \al_{nm}|c_m|^2 c_n \; + 
\frac{1}{2}\; \dlt_{n1}\bt_{12} c_2 e^{i\Dlt\cdot t} +
\frac{1}{2}\; \dlt_{n2}\bt_{12}^* c_1 e^{-i\Dlt\cdot t} \; ,
\ee
where
\be
\label{16}
\Dlt \equiv \Dlt\om +\al_{11}-\al_{22} \; .
\ee
For neighboring modes, one has $\al_{11}\approx\al_{22}$ and $\Dlt\approx
\Dlt\om$. Otherwise, it is always possible to choose such a detuning $\Dlt\om$
that $\Dlt$ be small, which is assumed in what follows, $|\Dlt|\ll\om$.

To preserve well-defined resonance, it is necessary, in analogy with the case 
of resonant atoms [35], that the transition amplitudes, involved in the 
process, be small. In the considered problem, such amplitudes are given by 
Eqs. (10) and (11). So, it is necessary that the transition amplitudes, 
due to atomic interactions, be small,
\be
\label{17}
\left | \frac{\al_{12}}{\om_{21}} \right | \ll 1 \; , \qquad
\left | \frac{\al_{21}}{\om_{21}} \right | \ll 1 \; ,
\ee
as well as the amplitude determined by the modulating field,
\be
\label{18}
\left | \frac{\bt_{12}}{\om_{21}} \right | \ll 1 \; .
\ee

From Eq. (15) it stems that if at the initial time $c_n(0)=0$ for $n\neq 1,2$,
then always
\be
\label{19}
c_n(t) = 0 \; , \qquad (n\neq 1,2)
\ee
for all $t\geq 0$. Hence Eq. (15) reduces to the system of two complex-valued 
equations
\be
\label{20}
i\; \frac{dc_1}{dt} = \al_{12} |c_2|^2 c_1 + \frac{1}{2}\; \bt_{12} c_2
e^{i\Dlt\cdot t} \; , \qquad
i\; \frac{dc_2}{dt} = \al_{21} |c_1|^2 c_2 + \frac{1}{2}\; \bt_{12}^* c_1
e^{-i\Dlt\cdot t} \; .
\ee
Such equations, though resembling the case of two coupled electromagnetic modes
[36], differ from that by the presence of the nonlinearity caused by atomic
interactions.

The complex-valued system (20) is equivalent to four real-valued equations.
However, because of the global gauge symmetry and due to the normalization
condition $|c_1|^2+|c_2|^2=1$, the related dynamical system is, actually,
two-dimensional. To show this, it is convenient to define $c_j=|c_j|
\exp(i\pi_j t)$ and $\bt_{12}\equiv\bt e^{i\gm}$, where $\bt\equiv|\bt_{12}|$. 
Let us also introduce the notation
$$
\al\equiv \frac{1}{2}\; (\al_{12}+\al_{21} ) \; , \qquad
\dlt \equiv \Dlt + \frac{1}{2}\; (\al_{12} -\al_{21} ) \; .
$$
Then for the population difference
\be
\label{21}
s \equiv |c_2|^2 - |c_1|^2
\ee
and effective phase difference
\be
\label{22}
x \equiv \pi_2 -\pi_1 +\gm +\Dlt \; ,
\ee
we find the system of two equations
\be
\label{23}
\frac{ds}{dt} = -\bt\sqrt{1-s^2}\; \sin x \; , \qquad
\frac{dx}{dt} = \al s + \frac{\bt s}{\sqrt{1-s^2}}\; \cos x + \dlt \; .
\ee

The consideration can be generalized to the case of the multiple generation 
of topological coherent modes [29,30]. This requires, instead of one modulating 
field (6), the action of several oscillating fields
\be
\label{24}
V(\br,t) = \frac{1}{2} \sum_j \left [ B_j(\br) e^{i\om_j t} +
B_j^*(\br) e^{-i\om_j t} \right ] \; ,
\ee
whose frequencies are tuned to the resonance with the chosen transition
frequencies. For instance, in the case of three coexisting modes, we obtain
$$
i\; \frac{dc_1}{dt} = \left ( a_{12}|c_2|^2 + \al_{13}|c_3|^2\right ) c_1 +
f_1 \; ,
$$
$$
i\; \frac{dc_2}{dt} = \left ( a_{21}|c_1|^2 + \al_{23}|c_3|^2\right ) c_2 +
f_2 \; ,
$$
\be
\label{25}
i\; \frac{dc_3}{dt} = \left ( a_{31}|c_1|^2 + \al_{32}|c_2|^2\right ) c_3 +
f_3 \; ,
\ee
where the forces $f_j$, related to the modulating fields, depend on the type
of the mode generation, whether this is the cascade generation, $V$-type, or
$\Lbd$-type generation [29,30]. The effective detuning
$\Dlt_{mn}\equiv\Dlt\om_{mn}+\al_{nn}-\al_{mm}$, where 
$\Dlt\om_{mn}\equiv\om_j-\om_{mn}$, is again assumed to be small,
$|\Dlt_{mn}|\ll\om_{mn}$.

In the same manner, one can derive the evolution equations for the amplitudes 
$c_n(t)$ for an arbitrary number of generated coherent modes, whose populations
are given by $|c_n|^2$. The resulting equations, such as (20), (23), or (25), 
are nonlinear because of the binary atomic interactions. One could also include
three-body interactions, which would yield the fifth-order nonlinearity with
respect to $|c_n|$. Such three-body interactions can play an important role in
describing the dissipation caused by three-body recombinations [1,37].

One could also take into account nonadiabatic corrections to the atomic 
evolution equations. However nonadiabatic description is of vital importance 
only for nonconfined motion of atoms [38], while for trapped atoms 
nonadiabatic corrections amount to at most a few percent [39].

\section{Dynamic Resonant Effects} 

Trapped condensed Bose gas, with resonantly generated topological coherent 
modes, resembles a resonant atom, hence, such a resonant condensate should 
display the features typical of resonant finite-level atoms. But this resonant
condensate is, in addition, a collective nonlinear system, because of which 
it can possess many other unusual properties, not existing in finite-level 
atoms.

\vskip 2mm

{\bf (1) Interference Fringes}.
The total density $\rho(\br,t)=|\vp(\br,t)|^2$ of trapped atoms, with generated
topological modes, is not simply the sum of the partial mode densities
$\rho_n(\br,t)=|c_n(t)\vp_n(\br)|^2$, but the interference fringes arise, 
described by the interference density
$$
\rho_{int}(\br,t) \equiv \rho(\br,t) - \sum_n \rho_n(\br,t) \; ,
$$
with fast oscillation in  time [20,23].

\vskip 2mm

{\bf (2) Interference Current}.
Similarly, there exists a fastly oscillating interference current
$$
{\bf j}_{int}(\br,t) \equiv {\bf j}(\br,t) - \sum_n {\bf j}_n(\br,t) \; ,
$$
sometimes called the internal Josephson current [20,23].

\vskip 2mm

{\bf (3) Mode Locking}.
Under this effect, the fractional mode populations $w_n$ are locked in the 
vicinity of their initial values, so that either
$$
0\leq w_n(t) \leq \frac{1}{2}
$$
for all $t\geq 0$, or
$$
\frac{1}{2} \leq w_n(t) \leq 1 \; ,
$$
never crossing the line $w_n=1/2$, but being either below it or above it, 
depending on initial conditions [6,23,24].

\vskip 2mm

{\bf (4) Dynamic Transition}.
Varying the system parameters, the dynamics of the mode populations can be
qualitatively changed from the mode locked regime to the mode unlocked regime, 
when the mode populations fluctuate in the whole region
$$
0 \leq w_n(t) \leq 1 \; ,
$$
independently of their initial values [15,23,24].

\vskip 2mm

{\bf (5) Critical Phenomena}.
On the manifold of the system parameters, there exists a critical surface in 
the vicinity of which the dynamics of the mode populations becomes unstable. 
Crossing the critical surface, when varying some parameters, changes the 
dynamics between the mode locked and mode unlocked regimes. Close to the 
surface, a tiny variation of some of the system parameters, say of the pumping
amplitude or of the detuning, provokes drastic changes in the dynamics of the 
mode populations [15,16,20,23,24]. The location of the critical surface also 
depends on the initial setup. Thus, for the case of two coexisting topological
modes, we have the critical surface described by the relation
$$
\frac{1}{2}\; \al s_0^2 - \bt\sqrt{1-s_0^2}\; \cos x_0 +\dlt s_0 = \bt\;
{\rm sgn}\;\al \; ,
$$
in which $s_0\equiv s(0)$ and $x_0\equiv x(0)$. For each given initial 
conditions, this is a surface in the three-dimensional space of the parameters
$\al,\bt$, and $\dlt$. For $s_0=\mp 1$, the critical surface reads as
$$
\bt\;{\rm sgn}\; \al \pm \dlt = \frac{1}{2}\; \al \; .
$$
Fixing $s_0=-1$, we can reduce the above relation to two critical lines on the
manifold of the parameters
$$
b \equiv \frac{\bt}{|\al|} \; , \qquad \varepsilon \equiv \frac{\dlt}{|\al|} 
\; .
$$
These lines are
$$
b+\varepsilon = \frac{1}{2} \qquad (\al > 0) \; ,
$$
$$
b -\varepsilon = \frac{1}{2} \qquad (\al < 0) \; .
$$
A time-averaged system displays on the critical surface critical phenomena 
typical of statistical systems with the second-order phase transitions. 
For the averaged system it is possible to define an effective capacity 
and susceptibility, which diverge on the critical surface [15,16,20].

\vskip 2mm

{\bf (6) Chaotic Motion}.
Fractional mode populations for a two-mode condensate are always periodic 
functions of time. But if the number of modes in the condensate is three 
or larger, then, depending on the system parameters, one has either 
quasiperiodic or chaotic motion. For instance, in the case of the 
three-mode condensate, for which $\al_{ij}=\al$, $\bt_{ij}=\bt$, and 
$\Dlt_{ij}=0$, chaotic motion appears when 
$$
\left | \frac{\bt}{\al} \right | \geq 0.639448 \; ,
$$ 
that is, under a sufficiently strong pumping [29,30].

\vskip 2mm

{\bf (7) Harmonic Generation}.
The generation of topological coherent modes may occur not solely under the 
direct resonance condition $\om=\om_{21}$, but also under the condition of 
harmonic generation
$$
n\om  =\om_{21} \qquad (n=1,2,\ldots) \; ,
$$
when just one modulating field, with a frequency $\om$, is applied [29,30].

\vskip 2mm

{\bf (8) Parametric Conversion}.
Another possibility of generating the topological modes is under the condition
of parametric conversion
$$
\sum_j (\pm \om_j) = \om_{21} \; ,
$$
when several alternating fields, with frequencies $\om_j$, are involved [29,30].

\vskip 2mm

{\bf (9) Atomic Squeezing}.
Squeezing is a quantum effect that does not exist for classical quantities. 
So, for treating it, one has to quantize the modes by considering the mode 
amplitudes $c_n$ as Bose operators. Then employing the pseudospin 
representation, one can define the squeezing factor
$$
Q \equiv \frac{2\Dlt^2(S_z)}{|<S_\pm>|} \; ,
$$
where $\Dlt^2(S_z)$ is the dispersion of the $z$-component of the total spin 
and $S_\pm$ are rising and lowering operators, respectively. The defined 
squeezing factor describes the relation between the dispersion, associated 
with the mode populations, and the dispersion of a relative phase. When 
$Q<1$, one says that  the atomic squeezing occurs. Then the mode populations 
can be measured with a higher accuracy as compared to the measurement of 
current [23].

\vskip 2mm

{\bf (10) Entanglement Production}.
In the condensate with topological coherent modes, entanglement can be 
produced. A general measure of entanglement production, valid for arbitrary 
systems, was introduced in Ref. [40]. The measure of entanglement, generated 
by an operator $\hat A$, is defined as
$$
\varepsilon(\hat A) \equiv \log\; 
\frac{||\hat A||_{\cal D}}{||\hat A^\otimes||_{\cal D}} \; ,
$$
where $||\cdot||_{\cal D}$ implies the restricted norm over a set ${\cal D}$ 
of disentangled functions, and $\hat A^\otimes$ is a nonentangling counterpart 
of $\hat A$. Considering a $p$-particle reduced density operator $\hat\rho_p$ 
for a multimode condensate, we find
$$
\varepsilon(\hat\rho_p) = (1-p)\log\sup_n\; w_n \; ,
$$
where $w_n=w_n(t)=|c_n(t)|^2$. Hence the above measure
$\varepsilon(\hat\rho_p)\equiv \varepsilon_p(t)$ is a function of time,
characterizing the entanglement evolution [40].

\vskip 5mm

Summarizing, we have shown that by applying resonant modulating fields to 
a Bose-condensed gas of trapped atoms, one can create topological coherent 
modes. We have developed an analytical theory describing the condensate with 
such coherent modes and also made numerical calculations by directly solving 
the Gross-Pitaevskii equation [29,30]. All results of the analytical theory 
are in very good agreement with numerical solutions. The condensate with 
topological coherent modes is a novel system possessing a rich variety of 
nontrivial properties which could be employed in many applications.

\vskip 5mm

{\bf Acknowledgement}

\vskip 2mm

One of the authors (V.I.Y.) is grateful to the German Research Foundation 
for the Mercator Professorship and another author (E.P.Y.) appreciates
financial support from the German Research Foundation under the DFG grant
Be 142/72-1.


\vskip 1cm

\begin{thebibliography}{99}

\bibitem{1}
P.W. Courteille, V.S. Bagnato, and V.I. Yukalov, Laser Phys. {\bf 11},
659(2001).

\bibitem{2}
A.L. Fetter and A.A. Svidzinsky, J. Phys. Condens. Matter {\bf 13}, 135 (2001).

\bibitem{3}
L. Pitaevskii and S. Stringari, {\it Bose-Einstein Condensation} (Clarendon,
Oxford, 2003).

\bibitem{4}
K. Bongs and K. Sengstock, Rep. Prog. Phys. {\bf 67}, 907 (2004).

\bibitem{5}
V.I. Yukalov, Laser Phys. Lett. {\bf 1}, 435 (2004).

\bibitem{6}
V.I. Yukalov, E.P. Yukalova, and V.S. Bagnato, Phys. Rev. A {\bf 56},
4845(1997).

\bibitem{7}
V.I. Yukalov, {\it Statistical Green's Functions} (Queen's University,
Kingston, 1998).

\bibitem{8}
K.P. Marzlin and W. Zhang, Phys. Rev. A {\bf 57}, 3801 (1998).

\bibitem{9}
K.P. Marzlin and W. Zhang, Phys. Rev. A {\bf 57}, 4761 (1998).

\bibitem{10}
E.A. Ostrovskaya, Y.S. Kivshar, M. Lisak, B. Hall, F. Cattani, and D. Anderson,
Phys. Rev. A {\bf 61}, 031601 (2000).

\bibitem{11}
D.L. Feder, M.S. Pindzola, L.A. Collins, B.I. Schneider, and S.W. Clark, Phys. 
Rev. A {\bf 62}, 053606 (2000).

\bibitem{12}
V.I. Yukalov, E.P. Yukalova, and V.S. Bagnato, Laser Phys. {\bf 10},
26 (2000).

\bibitem{13}
V.S. Bagnato, E.P. Yukalova, and V.I. Yukalov, in {\it Bose-Einstein
Condensates and Atom Lasers}, edited  by S. Martellucci, A. Chester, A. Aspect, 
and M. Inguscio  (Kluwer, New York, 2000), p. 201.

\bibitem{14}
Y.S. Kivshar, T.J. Alexander, and S.K. Turitsyn, Phys. Lett. A {\bf 278},
225 (2001).

\bibitem{15}
V.I. Yukalov, E.P. Yukalova, and V.S. Bagnato, Laser Phys. {\bf 11}, 455
(2001).

\bibitem{16}
V.I. Yukalov, E.P. Yukalova, and V.S. Bagnato, Proc. Int. Soc. Opt. Eng.
{\bf 4243}, 150 (2001).

\bibitem{17}
R.D' Agosta, B.A. Malomed, and C. Presilla, Laser Phys. {\bf 12}, 37 (2002).

\bibitem{18}
R.D' Agosta and C. Presilla, Phys. Rev. A {\bf 65}, 043609 (2002).

\bibitem{19}
N.P. Proukakis and L. Lambropoulos, Eur. Phys. J. D {\bf 19}, 355 (2002).

\bibitem{20}
V.I. Yukalov, E.P. Yukalova, and V.S. Bagnato, Laser Phys. {\bf 12},
231 (2002).

\bibitem{21}
V.I. Yukalov, E.P. Yukalova, and V.S. Bagnato, Laser Phys. {\bf 12},
1325 (2002).

\bibitem{22}
V.I. Yukalov, E.P. Yukalova, and V.S. Bagnato, Phys. Rev. A {\bf 66},
025602(2002).

\bibitem{23}
V.I. Yukalov, E.P. Yukalova, and V.S. Bagnato, Phys. Rev. A {\bf 66},
043602 (2002).

\bibitem{24}
V.I. Yukalov and E.P. Yukalova, J. Phys. A {\bf 35}, 8603 (2002).

\bibitem{25}
V.V. Serov, V.L. Derbov, S.I. Vinitsky, and V.I. Yukalov, Proc. Int. Soc. Opt.
Eng. {\bf 4706}, 124 (2002).

\bibitem{26}
V.I. Yukalov, E.P. Yukalova, and V.S. Bagnato, Laser Phys. {\bf 13},
551 (2003).

\bibitem{27}
V.I. Yukalov, E.P. Yukalova, and V.S. Bagnato, Laser Phys. {\bf 13},
861 (2003).

\bibitem{28}
J. Williams, R. Walser, J. Cooper, E.A. Cornell, and M. Holland, Phys. Rev. A
{\bf 61}, 033612 (2000).

\bibitem{29}
V.I. Yukalov, K.P. Marzlin, and E.P. Yukalova, Laser Phys. {\bf 14},
565 (2004).

\bibitem{30}
V.I. Yukalov, K.P. Marzlin, and E.P. Yukalova, Phys. Rev. A {\bf 69},
023620 (2004).

\bibitem{31}
S.K. Adhikari, Phys. Lett. A {\bf 308}, 302 (2003).

\bibitem{32}
S.K. Adhikari, J. Phys. B {\bf 36}, 1109 (2003).

\bibitem{33}
S.K. Adhikari, Phys. Rev. A {\bf 69}, 063613 (2004).

\bibitem{34}
N.N. Bogolubov and Y.A. Mitropolsky, {\it Asymptotic Methods in the Theory
of Nonlinear Oscillations} (Gordon and Breach, New York, 1961).

\bibitem{35}
L. Mandell and E. Wolf, {\it Optical Coherence and Quantum Optics} (Cambridge
University, Cambridge, 1995).

\bibitem{36}
A.S.M. de Castro, V.V. Dodonov, and S.S. Mizrahi, J. Opt. B {\bf 4}, 191 (2002).

\bibitem{37}
V.S. Filho, L. Tomio, A. Gammal, and T. Frederico, Phys. Lett. A {\bf 325}, 420
(2004).

\bibitem{38}
V.I. Yukalov, Phys. Rev. A {\bf 56}, 5004 (1997).

\bibitem{39}
J.H. M\"uller, O. Morsch, D. Ciampini, M. Anderlini, R. Mannella, and E. Arimondo,
Phys. Rev. Lett. {\bf 85}, 4454 (2000).

\bibitem{40}
V.I. Yukalov, Phys. Rev. A {\bf 68}, 022109 (2003).

\end{thebibliography}
\end{document}